\documentclass[twocolumn,showpacs,preprintnumbers,amsmath,amssymb]{revtex4}
\usepackage{graphicx}
\usepackage{dcolumn}
\usepackage{bm}

\begin{document}

\preprint{}

\title{Large oscillating non-local voltage in multi-terminal single wall carbon nanotube devices}
\author{G. Gunnarsson}
\author{J. Trbovic}
\author{C. Sch\"onenberger}
\affiliation{Department of Physics, University of Basel, Klingelbergstrasse 82, CH-4056 Basel, Switzerland}
\begin{abstract}
We report on the observation of a non-local voltage in a ballistic
(quasi) one-dimensional conductor, realized by a single-wall carbon
nanotube with four contacts. The contacts divide the tube into
three quantum dots which we control by the back-gate voltage
$V_g$. We measure a large \emph{oscillating} non-local voltage
$V_{nl}$ as a function of $V_g$. Though a classical resistor model
can account for a non-local voltage including change of sign, it
fails to describe the magnitude properly. The large amplitude of
$V_{nl}$ is due to quantum interference effects and can be
understood within the scattering-approach of electron transport.
\end{abstract}



\maketitle

The recent realization of the spin field-effect transistor in carbon
nanotube (CNT) devices~\cite{SahooNP}
demonstrated the ability to control spin transport in a quantum dot (QD)~\cite{CNT_Qdot}.
However, additional effects, such as the anomalous magneto\-resistance,
can contribute to the observed signal in spin-valves~\cite{Review2002,Roukes2003,Molenkamp2004,vdMolen2006}.
It seems clear, that despite a number of large
responses seen in CNT-based devices~\cite{SahooNP,Morpurgo2006,ZhaoAPL02,HuesoNat07},
one needs to go beyond two terminal structures by realizing multi-terminal
devices where non-local measurements are feasible~\cite{TombrosPRB06}.
The non-local measurement in spin-valve devices
has been pioneered by Johnson and Silsbee~\cite{Johnson85}
in metallic spin-valves and was further applied to various other systems~\cite{Jedema01,LouNP07,TombrosN07}.
%
This technique separates spin from charge effects.
Recent application of the \emph{non-local} spin technique in CNTs~\cite{TombrosPRB06} showed
the feasibility and yet tremendous challenge of performing such
measurements in low dimensional mesoscopic systems.
The hallmark of these measurements is that a
positive voltage is measured when the magnetization of the
injector and detector electrodes are parallel and a negative
\textit{only} when they are antiparallel. However, it has been
reported recently that the four-probe resistance with non-magnetic
probes in CNTs can be negative due to interference
effects~\cite{GaoPRL05}. This suggests that the measurement of the
\emph{non-local} spin transport in mesoscopic systems like CNTs
with ferromagnetic contacts should be strongly influenced by
quantum interference effects.

\begin{figure}
\includegraphics[width=8cm]{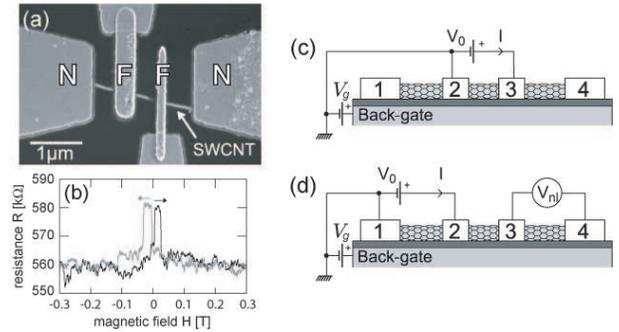}  
\caption{(a) An SEM-image of a device. The metallic
single-wall CNT is contacted with two ferromagnetic (F)
and two normal contacts (N), which together divide the tube into three
equidistant segments (\mbox{$L\approx 500$\,nm}) which act as QDs.
(b) Two terminal magneto-resistance signal (F-F).
(c,d) Schematics of the measurement setup for the
local two-terminal (c) and non-local four-terminal (d) measurement.}
\label{Fig1}
\end{figure}

We report here on measurements of a large
non-local voltage $V_{nl}$ in multi-terminal CNT devices (Fig.~1a)
in the quantum-dot (QD) regime
which changes sign and magnitude as the back-gate voltage is swept.
We show that $V_{nl}$ cannot be explained by
a classical resistor model. Instead, a quantum approach is required.
We also show that in these devices, which have relative transparent contacts with
resistances in the range of \mbox{$10-100$\,k$\Omega$},
the magnitude of the oscillating $V_{nl}$ greatly exceeds any
non-local spin signal.


Our devices consist of single-wall CNTs grown by chemical vapor deposition
(CVD) and con\-tac\-ted with four probes as shown in Fig 1a. Two middle
electrodes are ferro\-magnetic (F) made of
PdNi(20nm)/\,Co(25nm)/\,Pd(10nm) tri-layer, whereas the two outer probes
are normal (N) Pd(40nm) electrodes.  A PdNi alloy
with $30\%$ Pd is used, because it makes stable contacts to the CNT~\cite{SahooNP},
while Co serves as magnetization alignment layer for PdNi~\cite{Futurepub}.
The device lies on a $400$~nm thick $SiO_2$ layer, with an
underlaying highly doped Si substrate which is used as a back-gate.
The CNT was localized with a scanning-electron microscope (SEM) and the
structure was defined using electron-beam lithography.

\begin{figure}
\begin{center}
\includegraphics[width=8cm]{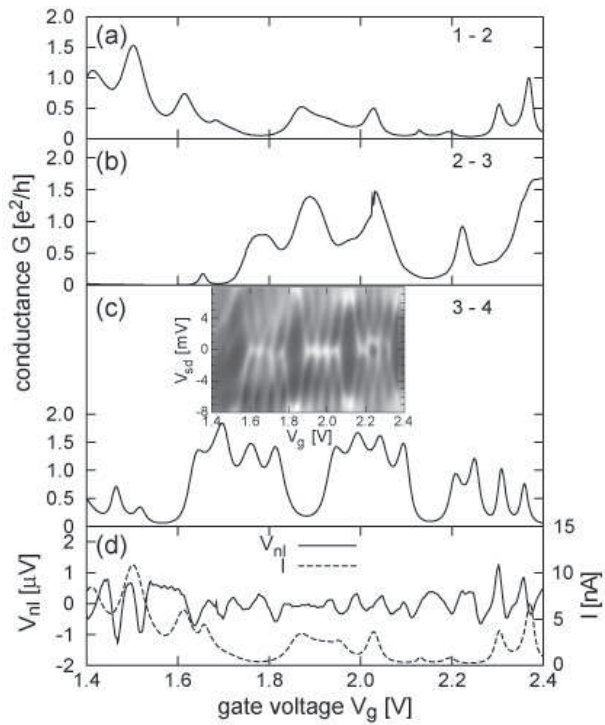}  
\caption{
  (a-c)Linear conductance $G$ of each segment of the
  device as a function of gate voltage $V_g$.
  The inset of \textbf{c} shows a gray-scale plot of $dI/dV_0$ as function of $V_g$ and
  source-drain bias $V_{sd}$ in the same $V_g$ range.
  (d) Non-local voltage measured across terminals 3 and 4 (full curve) and the
  current injected across terminals 1 and 2 (dashed).} \label{Fig2}
\end{center}
\end{figure}

Samples were cooled in a He4 cryostat to \mbox{$1.8$\,K} where the differential
conductance ($G=dI/dV_0$) was measured using standard low frequency
lock-in technique with an excitation voltage of \mbox{$V_0=100$\,$\mu$V}.
Two terminal local $G$-measurements were made across the three
segments of the sample (schematics in Fig~1c) in the gate
voltage range $V_g = 1.4-2.4$\,V, see Fig.~2a-c.
$G$ is found in the range of $0.1$ to \mbox{$2$\,$e^2/h$} and strongly
varies as a function of $V_g$, as expected
for a QD. By sweeping a DC source-drain voltage $V_{sd}$
we have obtained a  gray-scale plot of $G$ for the right segment $3-4$
(inset of Fig~2c).
%
%
The conductances of the three segments display qualitatively similar patterns
in different $V_g$-ranges, but the fine structure may vary.
This particular $V_g$-range has been selected
for Fig.~2, because of the pronounced four-fold pattern~\cite{BuitelaarPRL02,LiangPRL2002}
in  the detection arm of the non-local measurement, proving the absence of
intra\-tube scattering, and the presence of a QD formed by an \emph{individual}
single-wall CNT.

\begin{figure}
\begin{center}
\includegraphics[width=7cm]{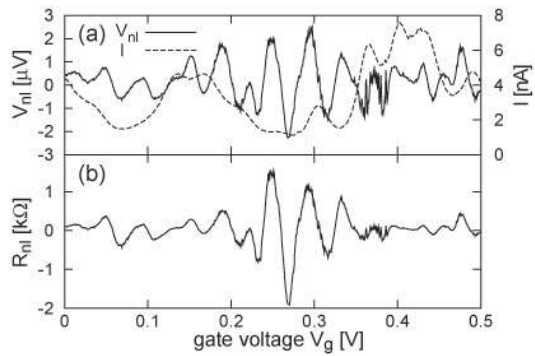} 
\caption{Non-local measurement as function of gate voltage $V_g$
of another device. (a) The current $I$ between terminals 1-2
(dashed) is plotted together with the non-local voltage $V_{nl}$ between
terminals 3-4 (solid). (b) Calculated non-local resistance $R=V_{nl}/I$.}
\label{Fig3}
\end{center}
\end{figure}

The non-local voltage measured across segment \mbox{3-4}
is shown in Fig~2d (full) together with the current $I$ (dashed curve)
injected through segment \mbox{1-2} and driven by a constant ac voltage $V_0$
of $200$~$\mu V$.
Two most striking features are noticeable in
these measurements. First, the non-local voltage $V_{nl}$ oscillates around zero,
and secondly, the amplitude is large with a typical
value of \mbox{$1$\,$\mu$V}. This results in an oscillating non-local resistance $R_{nl}=V_{nl}/I$
with values of \mbox{$0.1-1$\,k$\Omega$}. A similar behavior is found
in different gate-voltage ranges, as well as in different samples,
see e.g. Fig.~3.

In order to understand the origin of the observed signal, we first
model our device as a classical network of resistors, shown
in Fig.~4a~\cite{TombrosPRB06}.
Each terminal in the circuit is characterized by contact
resistances $R_{ci}$ and $r_{ci}$ and the CNT sections between terminals $i$ and its next
neighbors $j$ have resistances $R_{ij}$.
Two limiting cases are shown in (b) and (c) of Fig.~4:
in case (b) of `non-invasive' contacts ($R_c>>r_c$),
all contacts 1-4 couple weakly to the CNT.
In contrast, in case (c) of `strongly invasive' contacts ($R_c<<r_c$),
the CNT is split into segments. Although there have been reports on both
strong and weak contacts to CNTs~\cite{contacts-to-CNT},
a typical device lies in between.
%
For weak contacts (Fig.~4b), driving a current in the left branch results
in the appearance of a \emph{uniform} voltage $V'$ on the CNT.
Hence, $V_3=V_4=V'$ and the non-local voltage $V_{nl}:=V_3-V_4=0$.
For strong contacts  (Fig.~4c), $V_3$ and $V_4$ equal the bias voltage $V_0$, leading again
to a vanishing non-local voltage. Because we have assumed ideal
voltage probes in which no current flows, the right arm must have a uniform
potential also in the general case. We therefore conclude
that $V_{nl}=0$ in the classical limit.
This breaks down, if the elctro\-meters are not ideal, but possess a finite
input impedance, thereby providing a current sink.

\begin{figure}
\includegraphics[width=7cm]{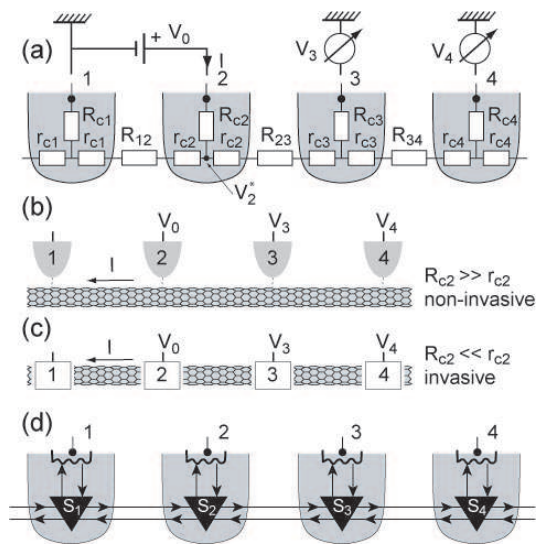} 
\caption{
(a) Resistor model for metal-CNT contacts.
The three resistors in each shaded region model the
property of the contact, whereas $R_{34}$, for example, describes the intra-tube
resistance in between contact $3$ and $4$.
Two limiting cases are shown in (b) and (c):
In (b) the contacts are weakly coupled to the CNT,
whereas they split the CNT into segments due to their strong coupling in (c).
(d) shows the quantum description, using three-port
scattering matrices $S_j$ (triangles) in each node.
Note, that there is an additional input resistance $R_I$ (not shown)
in both electro\-meters measuring $V_3$ and $V_4$.
} \label{Fig4}
\end{figure}

We next take the input amplifier input impedances \mbox{$R_I=100$\,M$\Omega$},
appearing at the voltage probes $3$ and $4$, in the resistor model in Fig.~4a
into account. Assuming a ballistic wire with $R_{ij}=0$, we obtain
for $V_{nl}$ of the third segment (3-4):
$V_{nl}\approx V_2^*(r_{c3}+r_{c4}+R_{c4}-R_{c3})/R_I$, 
where $V_2^*$ is the potential at the inner node of contact 2, as shown in
Fig.~4a. $V_2^*$ is of order $V_0$. Note, that because of the minus
sign in one term, a negative $V_{nl}$ is possible for certain resistance values.
If the magnitude of all contact resistances
were similar, however, a positive `mean' non-local voltage is predicted in
disagreement with the experiment. The observation of an oscillating $V_{nl}$
with a mean value close to zero could only be reconciled with the classical model
if $r_c << R_c$. Then, the equation simplifies and we arrive
at the following estimate: $V_{nl}\approx V_0 (R_{c4}-R_{c3})/R_I$.
This formula predicts that $V_{nl}$ follows the gate-voltage behavior of the
resistances of contact 3 and 4.
Using a typical contact resistance for our device of \mbox{$100$\,k$\Omega$},
\mbox{$R_I=100$\,M$\Omega$}, and \mbox{$V_0=200$\,$\mu$V}, we estimate
\mbox{$V_{nl}\sim 0.2$\,$\mu$V}. This is an order of magnitude smaller
than measured in the experiment, where the oscillating non-local voltage peaks up
to \mbox{$V_{nl}\sim 2$\,$\mu$V}. In order to be absolute certain that the current
in the detector arm caused by the finite input impedance $R_I$ of the amplifiers
is not the source of $V_{nl}$, we have crossed-checked these measurement
with a custom-modified amplifier with an input impedance of \mbox{$R_I=1$\,G$\Omega$}
and have found no change in $V_{nl}$. We are therefore confident that the
classical resistor model cannot account for the measured oscillating non-local
voltage $V_{nl}$ and that the finite input impedance of the amplifiers
is not the source of this signal.

\begin{figure}
\begin{center}
\includegraphics[width=7cm]{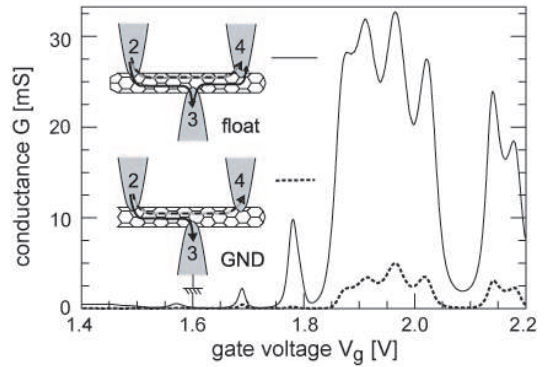} 
\caption{Measurements of the two-terminal conductance $G_{42}$
between terminals 2 and 4 while terminal 3 is floating (line) and
while it is grounded (dashed line). The inset depicts transmission through the CNT in both
cases.} \label{Fig5}
 \end{center}
\end{figure}

To understand the magnitude of the non-local voltage,
we next move on to a quantum-coherent description.
We apply the scattering approach~\cite{LB}
to calculate $V_{nl}$. A similar approach has been taken by Lerescu {\it et al.},
but for an open cavity that supports many modes~\cite{Lerescu}.
In contrast, in our case of a CNT we deal with a (quasi) one-dimensional (1d) system.
This approach is sketched in Fig.~4d. The triangles in each
node $j$ denote a three-port scattering matrix $S_j$. If we assume that each port is
described by a one mode conductor, $S_j$ is a $6\times 6$ matrix that describes
the amplitudes between outgoing and incoming waves~\cite{LB}. If we stick to
a one-mode conductor, one can prove that $V_{nl}=0$, independent of any details
of $S_j$. This theorem of vanishing non-local voltage in a 1d conductor can be
traced back to the particular structure of the whole $S$ matrix in which the $S_j$'s are
connected \emph{in series}. Hence, similar to the resistor model, even in the
coherent description, the non-local voltage is expected to disappear, provided
the wire is 1d. Because we do measure a large $V_{nl}$ in our devices,
one of the assumption in the theorem must be violated. These are: linear response and truly 1d.
To the former, we note that the maximum applied voltage of \mbox{$200$\,$\mu$V} corresponds
to \mbox{$2.3$\,K} which is slightly above the measurement temperature.
But we have also measured at \mbox{$100$\,$\mu$V} and below
confirming linearity. To the latter we emphasize that $V_{nl}$ does not change
with the relative magnetization direction of the two ferromagnetic contacts,
so that the spin degeneracy is not lifted. What remains as an explanation is the
fact that due to the so-called $K$ and $K^{\prime}$ degeneracy of graphene~\cite{K-Kprime}
any CNT should carry two (orbitally) degenerate 1d modes. This then leads to the
well-known four-fold pattern~\cite{BuitelaarPRL02,LiangPRL2002} in the spectrum of a CNT-QD,
which is clearly visible in Fig.~2c.

In order to estimate $V_{nl}$, we use the Landauer-B\"{u}ttiker formalism~\cite{LB}.
The current $I_i$ in lead $i$ is given by
$I_i \propto (N-R_{i})V_i-\sum T_{ij}V_j$, where $N$ denotes the number of modes (here $N=2$),
$R_i$ the total reflection coefficient for contact $i$,
$T_{ij}$ the probability that a charge carrier is transmitted from contact $j$ to contact $i$,
and $V_k$ the potential at contact $k$.
Using this formalism, one can derive a compact formula for a general four terminal voltage~\cite{LB}.
For our geometry
one obtains $V_{nl}=V_0(T_{32}T_{41}-T_{42}T_{31})/D$, where the denominator $D$ is given by:
$D=(N-R_4)(N-R_3)-T_{34}T_{43}$. This expression can only be evaluated (estimated) if we can,
in addition to nearest neighbor transmissions, estimate the transmission probabilities
that embrace second or even third-nearest neighbor contacts. These are the coefficients
$T_{31}$, $T_{42}$, and $T_{41}$. For the former two, the electron wave has to be able
to transmit \emph{under} one contact (2 or 3) without relaxation, while for the latter even
both contacts 2 and 3 need to be passed. We estimate this transmission beneath a contact
by comparing the conductance $G_{42}$ between terminals $2$ and $4$
when contact $3$ is floating or grounded. The result is shown in Fig.~5.
If the intermediate contact is floating, electrons that are transmitted into
contact 3 must be re-injected into the device so that the transmission between
2 and 4 can be large. On the other hand, for rather strong-coupling contacts as ours,
only a small fraction of electrons is expected to transmit unperturbed under the contact.
If contact 3 is now grounded, most carriers disappear via contact 3 to ground. This
latter situation is indeed realized as shown in the data of Fig.~5. The dashed curve
is suppressed by approximately a factor of $7-8$. Hence, a fraction of \mbox{$13$\,\%} can pass
under the contact by direct transmission~\cite{Makarovski}.
This is surprisingly large taken the typical
values of the two-terminal resistances and the fact that the metal electrodes were
evaporated directly onto a freshly CVD-grown CNT. Let us assume that transmission
under the contact is similar for contact 2 and 3 and let us denote this probability by $t$.
The magnitude of $V_{nl}/V_0$ is then estimated by $t^2$,
so that \mbox{$|V_{nl}|\alt 3.5$\,$\mu$V}, in good agreement
with the measured $V_{nl}$.

We now turn our attention to spin transport and estimate the
expected spin-dependent non-local voltage $V^{spin}_{nl}$.
We use the resistor model shown in Fig.~4. This time,
however, one has to expand it by introducing separate spin
up and spin down channels. The contact resistances at F contacts
depends on the relative orientation of the electron spin and the magnetization
within the F contacts~\cite{TombrosPRB06}. For simplicity we
assume that different contacts have equal resistances.
We emphasize here, that for the detection of a non-local spin
signal, the spin-imbalance in the injector part 1-2 has to be able
to `diffuse' into the detector branch. In the strongly invasive
limit this is impossible, because the strongly coupled
contact 2 would equilibrate the spin-imbalance.
In contrast, for weakly coupling contacts, a spin imbalance in the
CNT caused by spin injection can be sensed by a F contact.
$V^{spin}_{nl}$ therefore strongly depends on the ratio $r:=r_c/R_c$.
Assuming a small contact polarization $p$, one obtains
$V^{spin}_{nl}/V_0=p^2\cdot 0.25$, if $r << 1$, which is the largest
possible signal. Using the measurement shown in Fig.~5, we can estimate the $r$-ratio
for our devices. We obtain $r\simeq 6$. This then yields
$V^{spin}_{nl}/V_0 \simeq p^2\cdot 0.0037$.
We estimate $p$ from the typical two-terminal TMR of \mbox{$\approx 4$\,$\%$}~\cite{SahooNP},
yielding a polarization of \mbox{$\approx 20$\,$\%$}.
Inserting $p$ and \mbox{$V_0=200$\,$\mu$V}, the expected spin-dependent
non-local voltage is only \mbox{$V^{spin}_{nl}\simeq 30$\,nV},
two-orders of magnitude smaller than the measured non-local voltage.


In conclusion, we find large oscillating non-local signals in all of
our CNT quantum-dot devices. The existence of such a background, which changes
its sign with the back-gate voltage can screen the non-local signals
due to spin in particular in devices with relative high contact transparency.

{\it Note:} After completing this work, we have
become aware of a similar, but independent study by A. Makarovski {\it et al.}~\cite{Makarovski}.

{\it Acknowledgement:}
Support by the Swiss NSF, the NCCR on Nanoscale Science,
and EU-FP6-IST project HYSWITCH is gratefully acknowledged.
Fruitful discussions with B. J. van Wees are gratefully acknowledged.


\begin{thebibliography}{99}

\bibitem{SahooNP}
S. Sahoo, T. Kontos, J. Furer, C. Hoffmann, M. Gr{\"a}ber, A. Cottet, and C. Sch{\"o}nenberger,
Nature Phys. {\bf 1}, 99 (2005).
%
%
\bibitem{CNT_Qdot}
S. J. Tans, M. H. Devoret, H. Dai, A. Thess, R. E. Smalley, L. J. Georliga, and C. Dekker,
Nature {\bf 386}, 474 (1997);
M. Bockrath, D. H. Cobden, P. L. McEuen, N. G. Chopra, A. Zettl, A. Thess, and R. E. Smalley,
Science {\bf 275}, 1922 (1997);
D. H. Cobden, M. Bockrath, P. L. McEuen, A. G. Rinzler, and R. E. Smalley,
Phys. Rev. Lett. {\bf 81}, 681 (1998).
\bibitem{Review2002}
see for example:
H. X. Tang, F.G. Monzon, M. L. Roukes, F.J. Jedema, A.T. Filip and B.J. van Wees
in \emph{Semiconductor Spintronics and Quantum Computation},
D. D. Awschalom, N. Samarth and D. Loss eds., Springer Verlag (Berlin 2002).
\bibitem{Roukes2003}
H. X. Tang, R. K. Kawakami, D. D. Awschalom, and M. L. Roukes,
Phys. Rev. Lett. {\bf 90}, 107201 (2003).
\bibitem{Molenkamp2004}
C. Gould, C. R{\"u}ster, T. Jungwirth, E. Girgis, G. M. Schott, R. Giraud, K. Brunner,
G. Schmidt, and L.W. Molenkamp, Phys. Rev. Lett. {\bf 93}, 117203 (2004).
\bibitem{vdMolen2006}
S. J. van der Molen, N. Tombros, and B. J. van Wees, Phys. Rev. B {\bf 73}, 220406 (2006).
%
\bibitem{Morpurgo2006}
H. T. Man, I. J. W. Wever, A. F. Morpurgo, Phys.\ Rev.\ B {\bf 73}, 241401 (2006).
%
\bibitem{ZhaoAPL02}
B. Zhao, I. M{\"o}nch, H. Vinzelberg, T. M{\"u}hl, and C. M. Schneider, Appl.\ Phys.\ Lett.{\bf 80}, 3144 (2002).
%
\bibitem{HuesoNat07}
L. E. Hueso, J. M. Pruneda, V. Ferrari, G. Burnell, J. P. Valdes-Herrera, B. D. Simons, P. B. Littlewood,
E. Artacho, A. Fert, and N. D. Mathur, Nature {\bf 445}, 410 (2007).
%
\bibitem{TombrosPRB06}
N. Tombros, S. J. van der Molen, and B. J. van Wees, Phys.\ Rev.\ B {\bf 73}, 233403 (2006).
%
\bibitem{Johnson85}
M. Johnson and R. H. Silsbee, Phys.\ Rev.\ Lett. {\bf 55}, 1790 (1985).
%
\bibitem{Jedema01}
F. J. Jedema, A. T. Filip, B. J. van Wees, Nature {\bf 410}, 345 (2001).
%
\bibitem{LouNP07}
X. Lou, C. Adelmann, S. A. Crooker, E. S. Garlid, J. Zhang, K. S. M. Reddy, S. D. Flexner,
C. J. Palmstrom, and P. A. Crowell, Nature\ Phys.{\bf 3}, 197 (2007).
%
\bibitem{TombrosN07}
N. Tombros, C. Jozsa, M. Popinciuc, H. T. Jonkman, B. J. van Wees, Nature {\bf 448}, 571 (2007).
%
\bibitem{GaoPRL05}
B. Gao, Y. F. Chen, M. S. Fuhrer, D. C. Glattli, and A. Bachtold,
Phys.\ Rev.\ Lett. {\bf 95}, 196802 (2005).
%
\bibitem{Futurepub}
Detailed measurements of the magnetic properties of PdNi/Co bi-layer will be published elsewhere.
%
\bibitem{LiangPRL2002}
W. Liang, M. Bockrath, and H. Park, Phys. Rev. Lett. {\bf 88}, 126801 (2002).
%
\bibitem{BuitelaarPRL02}
M. R. Buitelaar, A. Bachtold, T. Nussbaumer, M. Iqbal, and C. Sch{\"o}nenberger,
Phys.\ Rev.\ Lett. {\bf 88}, 156801 (2002).
\bibitem{contacts-to-CNT}
see for example:
M. Bockrath, D. H. Cobden, L. Jia, A. G. Rinzler, R. E. Smalley, L. Balents, P. L. McEuen,
Nature {\bf 397}, 598 (1999);
A. Bachtold, M. de Jonge, K. Grove-Rasmussen, P. L. McEuen, M. Buitelaar, and C. Sch{\"o}nenberger,
Phys. Rev. Lett. {\bf 87}, 166801 (2001).
\bibitem{LB}
M. B\"{u}ttiker, Phys.\ Rev.\ Lett. {\bf 57}, 1761 (1986).
%
\bibitem{Lerescu}
A. I. Lerescu, E. J. Koop, C. H. van der Wal, B. J. van Wees, and J. H. Bardarson, arXiv:0705.3179v1.
\bibitem{K-Kprime}
M. S. Dresselhaus, G. Dresselhaus, P. C. Eklund: {\em Science of Fullerenes and Carbon Nanotubes}
(Academic Press, New York 1996).
\bibitem{Makarovski}
A. Makarovski, A. Zhukov, J. Liu, and G. Finkelstein, arXiv:0709.2498v1.
%
\bibitem{Julliere1975}
M. Julli{\'e}re, Phys. Lett. {\bf 54A}, 225 (1975).
\bibitem{Zutic2004}
I. Zutic, J. Fabian, and S. Das Sarma, Rev. Mod. Phys. {\bf 76}, 323 (2004).
%


\end{thebibliography}
\end{document}